\documentclass[10pt, a4paper]{article}

\usepackage[final]{lrec2026} 
\usepackage{tikz}
\usepackage{amsmath}
\usetikzlibrary{shapes.geometric, shapes.symbols, arrows.meta, positioning, shadows}
\usepackage{tcolorbox} 
\usepackage{booktabs} 
\usepackage{tabularx} 
\usepackage{caption}  
\usepackage{geometry} 
\usepackage{hyperref}
\usepackage{xurl}
\usepackage{multirow}
\usepackage{graphicx}
\usepackage{subcaption}
\usepackage{enumitem}
\usepackage{float}
\usepackage[T1]{fontenc}
\usepackage{listings}
\usepackage{upquote}

\title{HybridCodeAuthorship: A Benchmark Dataset for Line-Level Code Authorship Detection}

\name{Luke Patterson, Li Wang, Adam Faulkner}

\address{Card Intelligence, Capital One \\
         \{luke.patterson, li.wang, adam.faulkner\}@capitalone.com\\}

\abstract{
Thanks to the rapid adoption of AI code assistants powered by large language models (LLMs), industry codebases are, increasingly, a hybrid of AI- and human-authored code. For risk management and productivity analysis purposes, it is crucial to enable fine-grained location detection of AI-generated code. To develop algorithms for this task, quality benchmarks are needed to assess performance. However, existing benchmarks tend to comprise academic, LeetCode-style problems and presume a code snippet is either completely human-authored or completely AI-authored, which is not reflective of the diverse intents and styles of industry codebases utilizing AI code assistants. To fill these gaps, we introduce HybridCodeAuthorship, a novel benchmark of Python code files with interleaved human- and AI-authored lines of code to simulate authentic utilization of AI code assistants. In this paper, we first present our dataset construction pipeline, which leverages CodeSearchNet, a massive collection of links to open sourced repositories on GitHub. We then benchmark the performance of two state-of-the-art AI-generated code detection algorithms at both the line- and chunk-level. Experimental results demonstrate that HybridCodeAuthorship is a challenging benchmark with a top-scoring algorithm, AIGCode Detector, obtaining a highest F1 score of 0.48 and 0.56 on chunk-level and line-level code detection tasks, respectively.
\\
\newline
}

\begin{document}
\maketitleabstract
\section{Introduction}

Recent advances in generative AI are fundamentally reshaping the landscape of software development. AI code assistants powered by cutting-edge large language models (LLMs) are being rapidly adopted by industry to enhance developers' productivity.\footnote{A recent survey of developers from StackOverflow notes ``76\% of all respondents are using or are planning to use AI tools in their development process this year, an increase from last year (70\%).'' - \url{https://survey.stackoverflow.co/2024/ai}} A number of studies have been published that attempt to quantify the benefits of AI-based code assistants such as GitHub Copilot, Amazon CodeWhisperer, Google Cider, and Cursor. While most of the studies observed that developers who used AI code assistants experienced a 20\% to 56\% productivity increase  \citep{deniz2023unleashing, cui2025effects, peng2023impact, chatterjee2024impact, paradis2025much}, one study has reported the opposite --- experienced open-source contributors were decelerated by AI code assistants by \textasciitilde 19\% \citep{becker2025measuring}. To provide a more fine-grained view of developer adoption of these tools, additional work has measured the acceptance rates of coding suggestions and lines of code generated by AI code assistants \citep{ziegler2024measuring, bakal2025experience}. 

With rapid industrial adoption of coding assistants, hybrid code authorship has been described as a new paradigm of software development, with contemporary codebases increasingly a hybrid of interleaved human- and AI-generated code. For example, a mixed-methods study investigated the practical use of AI coding assistants \citep{Sergeyuk2025}, and highlighted the numerous interactions and edits that users typically made to AI-generated code. Recognizing this new paradigm, commercial engineering management platforms are increasingly adopting AI impact measures.\footnote{For examples, see: \url{https://jellyfish.co/blog/ai-impact-framework/} and \url{https://www.faros.ai/blog/how-much-code-is-ai-generated}} 

AI-generated code detection approaches typically involve extending existing AI-generated text detection approaches to the domain of code. This is sensibly motivated since distributional differences between human- and AI-generated natural language text, as measured by perplexity, have also been observed in human- versus AI-generated code \citep{xu2024detecting}. In benchmarking code detection approaches, researchers have relied on academic, LeetCode-style problem sets that bear little stylistic resemblance to authentic code produced as part of a typical software development project. A more accurate measure of the effectiveness of these approaches requires a dataset that 1) more accurately reflects the range of styles and intents found in practical codebases and 2) accurately reflects how developers make use of AI code suggestions by interleaving AI- and human-authored code. To the best of our knowledge, no current AI-generated code detection benchmark datasets meet these two criteria. To fill this gap, we introduce HybridCodeAuthorship, a novel benchmark dataset of Python code files implementing common software engineering tasks with human- and AI-generated code interleaved throughout. 

Our contribution is two-fold: 
\begin{enumerate}
\item A benchmark dataset, HybridCodeAuthorship, composed of full code files with interleaved human-authored and AI-generated code.
\item Experimental results showing initial benchmark performance of adapting two state-of-the-art AI-generated code detection algorithms for both line- and chunk-level code detection using HybridCodeAuthorship.
\end{enumerate}

As part of the first contribution, we created HybridCodeAuthorship \footnote{The complete benchmark is available at \url{https://github.com/CapitalOne-Research/c1-hybrid-code-authorship}} using a data construction pipeline that leverages CodeSearchNet \citep{husain2019codesearchnet}.  CodeSearchNet was released by GitHub and Microsoft Research, and is a massive collection of links to open-source repositories on GitHub.   CodeSearchNet contains links to approximately 2 million code files believed to be human-authored in popular programming languages such as Python, Java, and Go. The current version of HybridCodeAuthorship is limited to Python code snippets with the other programming languages reserved for future extensions. We defined two phases for the pipeline: code testing and code interleaving. The code testing phase was developed to verify the correctness of both human-authored and AI-generated code files. During this phase, each human-authored file and its AI-generated counterpart were labeled to indicate whether they passed or failed the same unit tests, respectively. These labels are included in the dataset release and, in practice, will allow experimenters to filter out non-functional code if their experiment setting requires such filtering. 

The code interleaving phase consists of three steps: code identification and masking, code marking, and, finally, code generation. In this latter phase, a specific portion of lines in each human-authored code file was selected, masked, and marked by an LLM with comments that could be recognized by an LLM for code generation. The resulting AI-generated code file was also validated for correctness with respect to Python syntax. The final dataset, in contrast to other benchmark datasets for AI code detection, accurately reflects real-world adoption of AI code assistants in software development by interleaving human-authored and AI-generated code relative to various acceptance rates.

Our interleaving algorithm is directly informed by recent mixed-methods research \citep{Sergeyuk2025} which highlights the frequent edits and fine-grained interactions developers have with AI-generated suggestions. We ask LLMs to first select distinct, atomic parts of human-authored code files and replace them with descriptive summaries of their functionality. These summaries serve as proxies for user-provided prompts, reflecting a developer’s specific intent for what a code block should achieve within a larger codebase. This approach mimics the authentic interaction of a user guiding a coding assistant to generate code that fulfills a specific requirement while maintaining the surrounding human-authored context. To ensure the realism of the resulting hybrid code, we use complex, real-world repositories from CodeSearchNet as our foundation. We further validate the authenticity of these interleaved files by requiring them to pass the original project unit tests and maintain syntactical correctness, ensuring they represent functional software rather than just superficial code fragments.

Our second contribution, experimental results utilizing two state-of-the-art AI-generated code detection algorithms, demonstrate that line- and chunk-level code detection is a challenging task with a top-scoring approach, AIGCode Detector \citep{xu2024detecting}, obtaining a highest F1 score of 0.48 and 0.56 on chunk-level and line-level code detection tasks, respectively. Line-level code detection treats each line of code as a distinct unit for authorship classification. Chunk-level first creates ``chunks'' by concatenating consecutive lines of code that share the same author type, and treats those chunks as the units for classification. 

The rest of the paper is organized as follows: Section 2 reviews previous work on benchmark datasets and relevant algorithms for AI code detection. Section 3 describes the two core phases powering the data construction pipeline while Section 4 describes the resulting  benchmark dataset, HybridCodeAuthorship. Section 5 presents experiment results with two selected approaches to AI-generated code detection. Section 6 discusses the limitations of our work. Finally, Section 7 provides a summary of our work and outlines our plans for future research. We hope that our work will accelerate the development of new algorithms and inspire future studies that explore topics beyond validating productivity enhancement driven by AI code assistants.
\section{Related Work}

In recent years, numerous benchmark datasets have been proposed to evaluate the performance of AI code detection algorithms. \citet{Alam2023} developed GPTCloneBench, a large dataset of AI-Human clone pairs generated with GPT-3 \citep{brown2020language}. \citet{Demirok2024} created AIGCodeSet, containing human-authored and AI-generated pairs using a variety of LLMs. Several other datasets have been proposed in recent years \citep{Bulla2024, Idialu2024, Pan2024, Shi2025Between, xu2024detecting, Choi2025, demirok2025multiaigcd}. Recently, newer datasets have expanded the number of languages, models, and code files. \citet{guo2025codemirage} introduced a dataset of over 200,000 sample code files including 10 different languages and uses 10 different LLMs to generate code. Furthermore, \citet{orel2025droid} introduced an extensive resource suite including, DroidCollection, a large-scale dataset of over one million code samples across 7 programming languages, 43 language models and 3 coding domains. 

However, the construction of these datasets remains similar: the authors start with some set of prompts and human-authored solutions. Then, the authors ask an LLM or LLMs to produce AI-generated code for the same prompt. In almost all existing benchmark datasets, code snippets are either entirely human-authored or entirely AI-generated. These benchmarks are designed for the task of binary authorship classification on code snippets. \citet{orel2025droid} proposed machine-refined samples to simulate three coding scenarios of human-AI interaction in the real world: human-to-LLM continuation, gap filling and code rewriting. However, the machine-refined samples still target coarse-grained classification tasks with sample-level classes. To reflect the evolving trend of hybrid authorship in code, detection algorithms must be tested at a more finer-grained unit of analysis, such as the line-level.   

There are now numerous commercially available tools dedicated to the task of AI-generated text detection \citep{copyleaks, gptzero, openai, writer} and both classification \citep{guo2023close, solaiman2019release} and perturbation-based approaches \citep{mitchell2023detectgpt, xu2024detecting}  to the task have been proposed in the literature. Line-level code authorship attribution remains an under-researched task  with  a handful of approaches existing in the stylometry and authorship attribution literature predating contemporary LLMs. \citet{caliskan2015de} developed a random forest stylometric classifier with features derived from abstract syntax trees of code, \citet{wang2018integration} introduced an approach that uses features generated dynamically from code execution to detect authorship, and \citet{Abuhamad2020Multi} implemented an RNN-based authorship verification model for this task. Recently, \citet{orel2025droid} trained a suite of Transformer encoders, DroidDetect, using a supervised contrastive objective over DroidCollection after removing suspicious AI code. \citet{xu2024detecting} developed AIGCode Detector, which is an adaptation of DetectGPT. It identifies AI-generated code by comparing the log probabilities of original and perturbed versions, utilizing CodeBERT for perturbations and a composite scoring system based on perplexity, standard deviation, and burstiness.

This concept of hybrid authorship also applies to the domain of natural language text, and several benchmark datasets of text have already been constructed. \citet{kadiyala2025robust} produced a large-scale dataset with 2.4 million hybrid texts in 23 languages from 12 different models. Similarly, \citet{su2025haco} introduced HACo-Det, a dataset generated via an automatic pipeline that provides word-level attribution labels. Other notable datasets for detecting hybrid prose include MixSet \citep{zhang2024llm}, RAID \citep{dugan2024raid} and Beemo \citep{artemova2024beemo}. In this paper, we extend this process to the production of hybrid authorship Python code.
\section{Benchmark Construction}

\begin{figure*}[t]
\centering
\includegraphics[width=\textwidth]{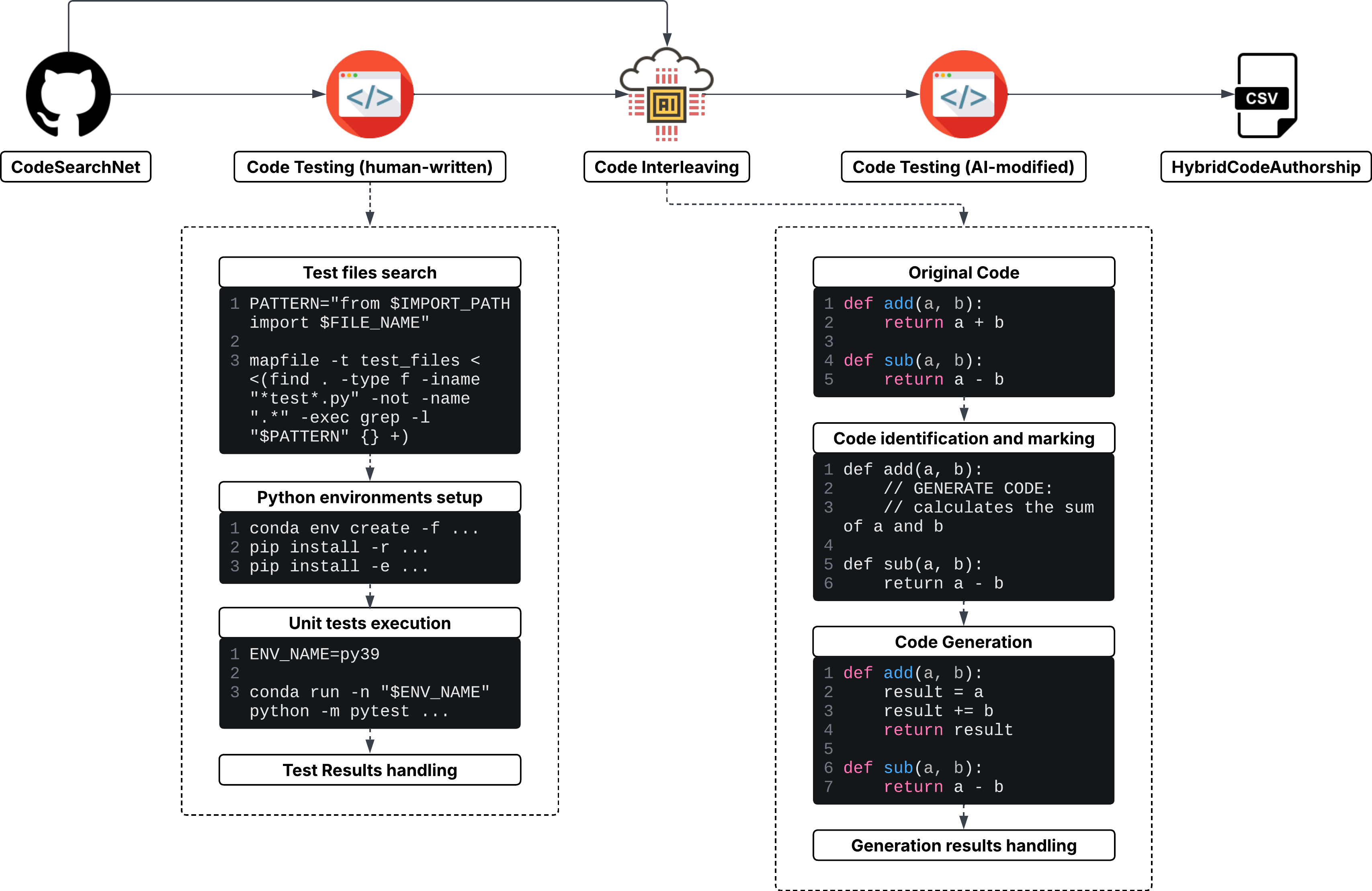}
\caption{Overview of the data construction pipeline for HybridCodeAuthorship. First, human-authored code files sampled from CodeSearchNet were validated during a code testing phase, which involves test files search, Python environment setup, and unit test execution. Second, all human-authored files were processed by the code interleaving phase in a parallel step that performed code identification, code marking (i.e., instructing an LLM to provide a comment for the downstream LLM to write code that fulfills the intent of the masked code) and, finally, code generation using an LLM. Third, the AI-modified code files were verified via another code testing phase. Finally, both human-authored and AI-modified code files were collected and annotated at the line-level as either ``Human'' or ``AI'', resulting in a final benchmark dataset.}
\label{fig:pipeline_overview}
\end{figure*}

We provide an illustration of the data construction pipeline for our benchmark dataset, HybridCodeAuthorship, in Figure \ref{fig:pipeline_overview}. The pipeline processes code files sampled from CodeSearchNet in two phases: code testing and code interleaving. Notably, both phases comprise multiple steps and code testing is invoked twice to run the same unit tests against each human-authored code file and its AI-modified counterpart.

\subsection{CodeSearchNet}

CodeSearchNet \citep{husain2019codesearchnet}, released by GitHub and Microsoft Research, is a massive collection of links to open-source repositories on GitHub. It contains approximately 2 million links to code files written in 6 programming languages (i.e., Go, Java, JavaScript, PHP, Python and Ruby). The Python subset of links in CodeSearchNet points to more than 450,000 Python code files from over 13,000 unique repositories with diverse levels of complexity and intent. Each file is identified by its metadata including GitHub URL, repository name, file path, function name, docstring, function code and commit hash. As CodeSearchNet was originally proposed to facilitate model training for semantic code search, the code files were further split into training, validation and test sets.

Released in 2019 before the emergence of LLMs, CodeSearchNet is generally believed to represent real-world code written by humans and is considered a standard benchmark for evaluating algorithms for code generation and detection \citep{feng2020codebert, wang2021codet5, lu2021codexglue, wang2023codet5+, li2023starcoder, xu2025distinguishing}. In addition to being human-authored, CodeSearchNet was selected as a representative sample of GitHub repositories implementing software solutions for real-world tasks. 

To construct HybridCodeAuthorship, we selected a subset of 4,814 Python code files from CodeSearchNet that are still available on GitHub.

\subsection{Code Testing}

As part of our broader goal of creating a dataset that reflects real-world coding tasks, we first assessed the correctness of the code as code correctness is generally viewed as a prerequisite of functioning software. To provide information about the validity of both human-authored and AI-modified code in HybridCodeAuthorship, we developed a code testing phase that verifies code files by running relevant unit tests from their repositories using \texttt{Bash}. The same steps were followed to test both human-authored and AI-modified code.

\subsubsection{Test Files Search}

Given a code file (either human-authored or AI-modified) and its repository, code testing started with a search for all test files that import the target code file within the associated commit and the \texttt{main}/\texttt{master} branch. Ancestor directories of the target file were first scanned for non-hidden folders containing the \texttt{test} handle. If one or more test folders were found, our code testing framework then searched for all test files in each test folder that contain common import statements mentioning the target code file using a regular expression:
\begin{lstlisting}[language=Python, basicstyle=\small\ttfamily, breaklines=true, columns=fullflexible]
'\b(import|from)\b[^#]*\b($PARENT_DIR|$FILE_NAME)\b'
\end{lstlisting}
Notably, \texttt{PARENT\_DIR} is included to handle the edge case where \texttt{FILE\_NAME} is \texttt{\_\_init\_\_.py}. For example, given a target file \url{https://github.com/keon/algorithms/blob/4d6569464a62a75c1357acc97e2dd32ee2f9f4a3/algorithms/queues/priority_queue.py}, all the file paths under \path{/algorithms/algorithms/queues/priority_queue.py} were scanned for common import statements like:
{
\scriptsize
\begin{itemize}
\item[]
\verb|import priority_queue|

\item[]
\verb|import queues.priority_queue|

\item[]
\verb|import algorithms.queues.priority_queue|

\item[]
\verb|from priority_queue import|

\item[]
\verb|from queues.priority_queue import|

\item[]
\verb|from algorithms.queues.priority_queue import|
\end{itemize}
}
This approach to selectively executing unit tests, rather than the entire suite, minimizes false positive results that occur when the code file passes tests that are unrelated to its functionality and boosts the efficiency of the entire pipeline.

\subsubsection{Python Environment Setup}

Following a successful search of eligible test files, code testing proceeded with multiple Python environments whose versions were tied to the code being tested. We made the following key decisions related to environment setup:
\begin{itemize}
\item
Python versions (2.7, 3.6 and 3.12) were aligned to the time range in which code files of CodeSearchNet were created.

\item
Compatible versions of common libraries, such as \texttt{numpy}, \texttt{scipy}, \texttt{mock}, etc., were pre-installed as part of the ``default'' environment.  

\item
The open-source testing framework, \texttt{pytest}, which contains built-in support of other testing libraries such as \texttt{unittest} and \texttt{nose},
was used to execute all unit tests.
\end{itemize} 
To minimize import errors during unit tests and mitigate false negative results, we also installed the repository and its required libraries through \texttt{setup.py}, \texttt{pyproject.toml} and \texttt{requirements.txt} files. Crucially, each repository was installed in a clean virtual environment to avoid dependency conflicts.

\subsubsection{Test Execution and Correction}

To evaluate the functional correctness of both human-authored and AI-modified code, we implemented an iterative testing methodology. This strategy systematically filtered out the code snippets that passed all tests in any eligible test file during a given run, allowing subsequent runs to focus on only those that failed all tests with upgraded testing code based on failure logs. Furthermore, the iterative filtering in the strategy prevents redundant test executions and optimizes the utilization of computation resources. 

With a large number of code files, it is practically infeasible to test each single function in a code file with the hope to pass all tests across multiple test files, without exaggerating the likelihood of false negative results. By adopting a greedy heuristic classifying a code snippet as a success if it passes all tests in any eligible test file, we believe we effectively balanced the trade-off between false negatives and false positives.
 
Although required dependencies were mostly captured by the environment setup stage, \texttt{pytest} could still raise \texttt{ModuleNotFound} error during run time. To further mitigate false negative results, we developed an adaptive correction mechanism that attempted to install missing libraries based on the error messages from \texttt{pytest} during a testing run. For libraries whose names cannot be inferred from their import statements (e.g., scikit-learn), we maintained a hard-coded list of missing libraries throughout the iterations.  

\subsubsection{Test Results Handling}

We added two columns \texttt{HumanCodeTier} and \texttt{AICodeTier} in HybridCodeAuthorship to indicate the validity of each human and AI code file, respectively. Each column comprises of three validity levels: ``Unit Test Passed'', ``AST parsable'' and ``Unparsable''. If a code file passed any test running within a Python environment and a repository state, it was annotated with ``Unit Test Passed''. Otherwise, if the code file was still AST (abstract syntax tree) parsable, it received the annotation ``AST parsable''. Because AST is dependent on Python version, we also included the version in the ``AST parsable'' label for completeness (e.g., ``AST parsable: py36'').

Unit tests associated with a target code file could fail for a variety of reasons, especially for legacy repositories. The most common reasons for both human-authored and AI-modified code failure were:
\begin{itemize}
\item
\texttt{Tests Not Found}: No test files matched the common import patterns or the test files were created in a fork (i.e., copy of the repository).

\item
\texttt{Import Error}: Required modules did not exist in available libraries or the import paths of custom modules were broken.

\item
\texttt{Module Not Found Error}: Required libraries were missing for unit tests or could not be installed.

\item
\texttt{Syntax Error}: Code in the target code file or test files violated Python grammar.

\item
\texttt{Repo Not Found}: Repositories became missing or inaccessible or could not be installed.
\end{itemize}
\noindent Notably, AI-modified code tended to produce new errors not observed with human-authored code, including \texttt{Attribute Error}, \texttt{Type Error}, \texttt{Name Error} and \texttt{Value Error}. Please refer to Appendix~\ref{appendix_a} for complete error analysis. 

\subsection{Code Interleaving}

We attempted code interleaving on a set of 4,814 human-authored code files. Segments in each file were identified, masked, and subsequently replaced with AI-generated code that preserved the original functionality and intent of the masked code. This process of \textit{AI code interleaving} was performed in three steps for each file. These steps were repeated for three different LLMs, \texttt{Llama3.3-70B}, \texttt{Llama-4-Scout} \citep{dubey2024llama}, and \texttt{GPT-OSS-120b} \citep{AgarwalEtAl2025}. These steps are visualized in Figure \ref{fig:pipeline_overview}. 4,196 files successfully completed our pipeline for at least one LLM.

\begin{table*}[h]
\centering
\captionsetup{skip=5pt}
\begin{tabularx}{\textwidth}{@{}llX@{}}
\toprule
\textbf{Column Name} & \textbf{Data Type} & \textbf{Description} \\
\midrule
\texttt{ModelId} & String & Identifier for which LLM was used for code generation. \\
\texttt{RecordId} & String & A unique identifier for each code sample, serving as the primary key when combined with \texttt{ModelId}. \\
\texttt{Language} & String & The programming language of the code sample. \\
\texttt{GitHubUrl} & String & The URL to the original GitHub repository and file. \\
\texttt{HumanCode} & String & The original, unmodified human-authored code. \\
\texttt{HumanCodeTier} & String & The validity tier of the human-authored code (``Unit Test Passed'', ``AST Parsable'', ``Unparsable''). \\
\texttt{AICode} & String & The final code containing interleaved AI-generated content. \\
\texttt{AICodeTier} & String & The validity tier of the AI-generated code (``Unit Test Passed'', ``AST Parsable'', ``Unparsable''). \\
\texttt{AICodeLines} & List of String & \texttt{AICode} string as a list with each line its own item. \\
\texttt{LineNumber} & List of Int & List of line numbers for \texttt{AICode}. Used as index for lists in \texttt{AICodeLines}, \texttt{Attribution}, and \texttt{Triviality} columns. \\
\texttt{Attribution} & List of String & List of attribution labels for \texttt{AICode} lines (``AI'' or ``Human''). \\ 
\texttt{Triviality} & List of String & List of triviality labels for \texttt{AICode} lines (``Trivial'' or ``Nontrivial''). \\
\texttt{AILineProportion} & Float & The actual proportion of lines attributed to AI in the \texttt{AICode}. \\
\bottomrule
\end{tabularx}
\caption{HybridCodeAuthorship schema.} 
\label{tab:main_schema}
\end{table*}

\subsubsection{Code Identification}

In the first, code identification step, the LLM was prompted to identify lines of meaningful, atomic code for replacement. To modulate the extent of code modification, the prompt included a target replacement percentage, denoted as $p$. This parameter was intended to control the density of AI-generated code within the output interleaved file. Over all code files, the value of $p$ was systematically varied, sampling uniformly from the discrete set $\{10\%, 20\%, \dots, 100\%\}$. While LLM outputs did not always adhere to this request, we empirically observed that inclusion of this instruction helped increase overall variance in AI code density among code files. We also left a random sample of files (10\%) completely unchanged and marked all lines as \texttt{Human}. This increased the difficulty of the AI-generated code detection task and improves the robustness of the resulting benchmark where code detection algorithms cannot rely on file-level signal potentially correlated with the existence of AI-generated line of code in the file. 

\subsubsection{Code Marking}

In the second, code marking step, the LLM was instructed to remove the original code identified lines in step 1 and to replace the code with a placeholder comment starting with the string \texttt{GENERATE CODE:}. This clearly identified the section of code to be generated in step 3. Additionally, the LLM was instructed to include a comment that  summarizes the intent of the deleted code in enough detail to allow plausible reconstruction with similar functionality, though not necessarily with the same exact syntax. See Appendix~\ref{appendix_b} for the code marking prompt.

\subsubsection{Code Generation}

In the final step, the marked source file was presented to the LLM for code generation. During this phase, the model was asked to parse the file, identify all instances of the \texttt{GENERATE CODE:} placeholders and, for each one, to generate a new code segment that met the requirements described in the associated comment. See Appendix~\ref{appendix_c} for the code generation prompt.   

\subsection{Determining Line-Level Authorship}

For each interleaved code file, we must reconcile which lines are AI-generated and which are human-authored. To do this, we used the Python \texttt{diff} library to track which lines were preserved from the original code (marked as \texttt{Human}) versus which lines were newly introduced in the interleaving process (marked as \texttt{AI}). This provides the ground-truth labeling needed for benchmark evaluation. 

To focus evaluation on substantive code portions, we also classified lines as either \texttt{Trivial} or \texttt{Nontrivial}. This allows experimenters to vary the segments targeted by their AI-generated code detection algorithms depending on the requirements of their experiments. We used \texttt{regex} matching logic to identify lines with minimal semantic content:
\begin{lstlisting}[language=Python, basicstyle=\small\ttfamily, breaklines=true, columns=fullflexible]
r'^[a-zA-Z_][a-zA-Z0-9_]*\s*=\s*(?:True|False|\d+|"[^"]*"|\'[^\']*\')$'
\end{lstlisting}
A line is also marked as \texttt{Trivial} if it consists of whitespace, simple comments, docstrings, \texttt{pass} statements, or standalone brackets. 
\section{HybridCodeAuthorship Dataset}

\begin{figure*}[t]
\centering 

\begin{subfigure}[t]{0.48\textwidth}
\centering
\includegraphics[width=\linewidth]{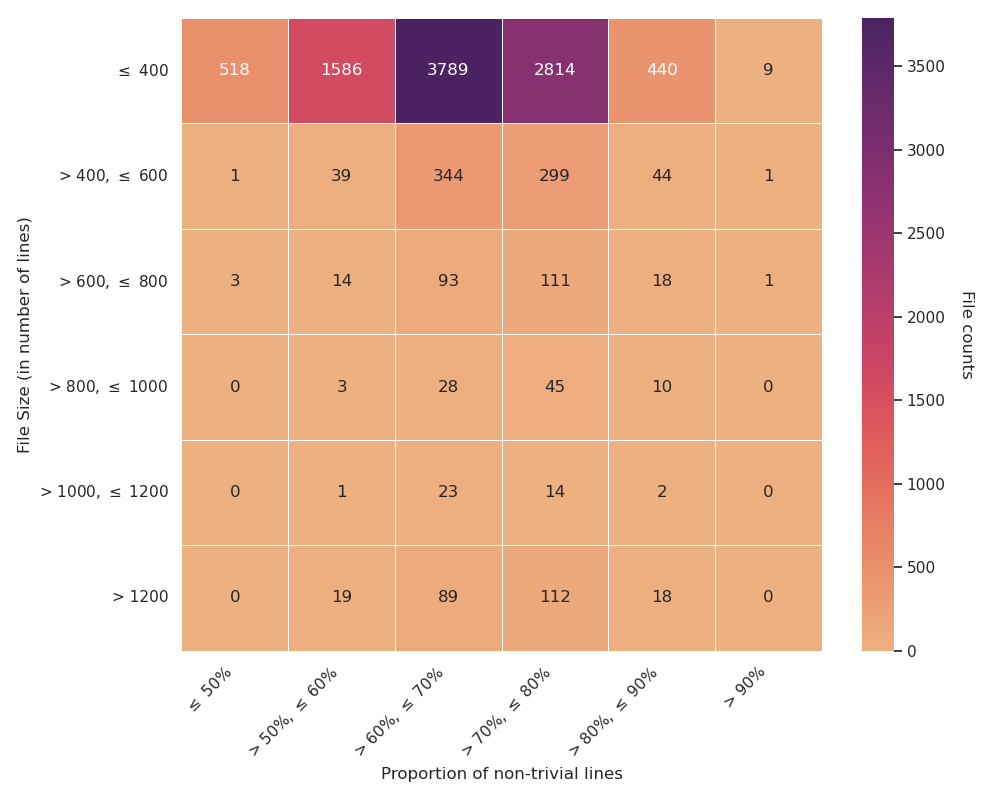}
\caption{Number of files by file size and proportion of nontrivial lines for the entire benchmark.}
\label{fig:heatmap_count1}
\end{subfigure}
\hfill 
\begin{subfigure}[t]{0.48\textwidth}
\centering
\includegraphics[width=\linewidth]{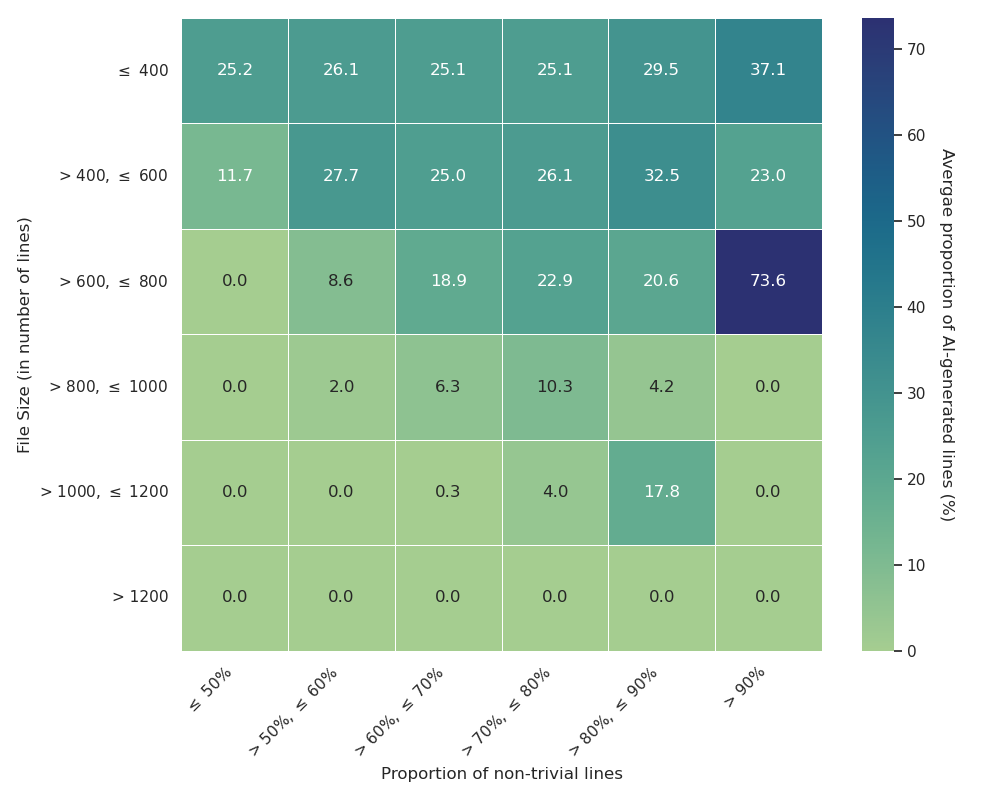}
\caption{Average proportion of AI-generated lines by file size and proportion of nontrivial lines for the entire benchmark.}
\label{fig:heatmap_ai1}
\end{subfigure}

    
\begin{subfigure}[t]{0.48\textwidth}
\centering
\includegraphics[width=\linewidth]{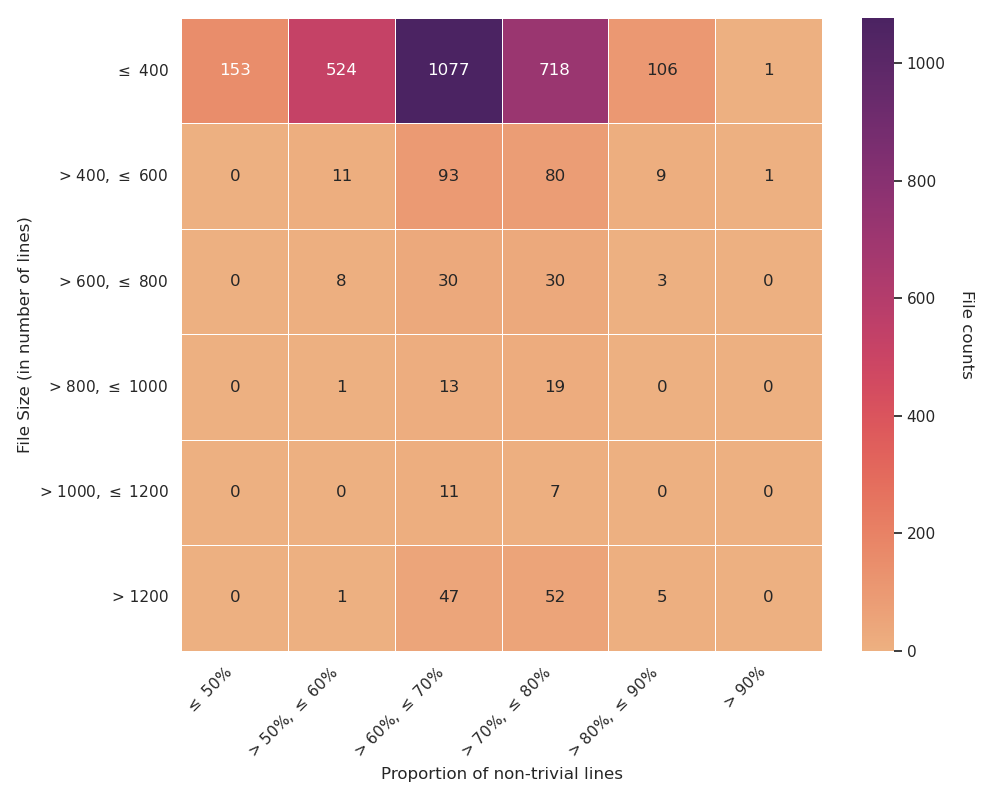}
\caption{Number of files by file size and proportion of nontrivial lines for the subset where both human-authored and AI-generated files passed unit tests.}
\label{fig:heatmap_count2}
\end{subfigure}
\hfill 
\begin{subfigure}[t]{0.48\textwidth}
\centering
\includegraphics[width=\linewidth]{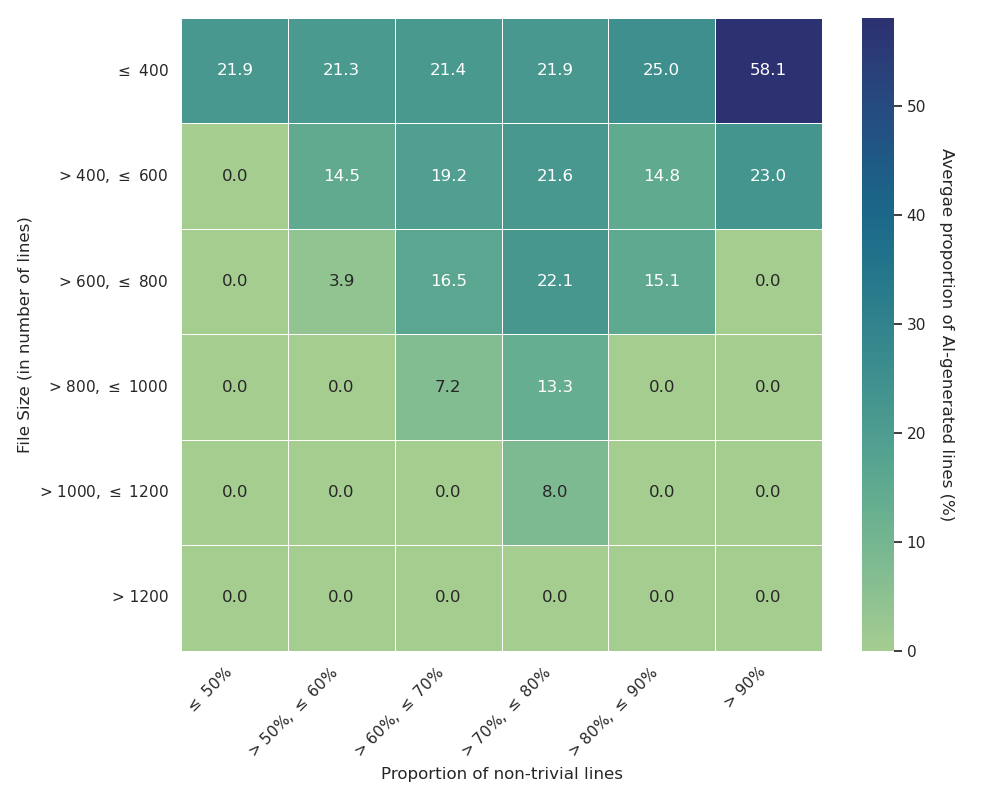}
\caption{Average proportion of AI-generated lines by file size and proportion of nontrivial lines for the subset where both human-authored and AI-generated files passed unit tests.}
\label{fig:heatmap_ai2}
\end{subfigure}

\caption{Overview of HybridCodeAuthorship with respect to the distributions of file counts and proportion of AI-generated lines over file size and proportion of nontrivial lines. HybridCodeAuthorship demonstrates sufficient variation in data segments with high concentration in code files that have fewer than 1000 total lines and over 50\% nontrivial lines.}
\label{fig:benchmark}
\end{figure*}

The HybridCodeAuthorship dataset comprises 10,488 records derived from 4,196 Python code files. Each file was independently rewritten by multiple LLMs, resulting in several modified versions per source file. Of the 10,488 file records, 39\% (4,103) of human-authored code files passed unit tests. In comparison, 29\% (3,000) of the 10,488 AI-interleaved records passed unit tests. Within these files, there is a total of 2,827,938 lines of code. 17\% (488,896) of lines are AI-generated, while the rest are human-authored. 69\% (1,943,728) of lines are deemed nontrivial. The schema for the dataset, detailing its columns and data types, is presented in Table \ref{tab:main_schema}. The distributions of file counts and average proportions of AI-generated lines across file sizes and proportions of nontrivial lines are visualized in Figure \ref{fig:benchmark}. The top two plots overview the entire benchmark and the bottom two zoom into the subset where both human-authored and AI-generated code files passed unit tests. Regarding data utilization, we recommend taking additional data processing steps to balance the proportions of AI-generated lines across files with different sizes. 

\section{Experimental Results}

\begin{table*}[t]
\centering
\begin{tabular}{cccccc}
\toprule
\textbf{LLM} & \textbf{Dataset Split} & \textbf{Granularity} & \textbf{Metric} & \multicolumn{2}{c}{\textbf{Detection Algorithm}} \\
\cmidrule{5-6}
& & & & \textbf{DroidDetect} & \textbf{AIGCode Detector} \\
\midrule
\multirow{12}{*}{GPT-OSS-120b} & \multirow{6}{*}{Trivial} & \multirow{3}{*}{Line-level} & Precision & $0.197$ & $0.560$ \\
& & & Recall & $0.986$ & $0.570 $ \\
& & & F1 & $0.328$ & $\textbf{0.560}$ \\
\cmidrule{3-6}. 
& & \multirow{3}{*}{Chunk-level} & Precision & $0.101$ & $0.540$ \\
& & & Recall & $0.856$ & $0.560 $ \\
& & & F1 & $0.181$ & $\textbf{0.480}$ \\
\cmidrule{2-6}. 
& \multirow{6}{*}{Nontrivial} & \multirow{3}{*}{Line-level} & Precision & $0.266$ & $0.540$ \\
& & & Recall & $0.994$ & $0.550 $ \\
& & & F1 & $0.419$ & $\textbf{0.530}$ \\
\cmidrule{3-6} 
& & \multirow{3}{*}{Chunk-level} & Precision & $0.144$ & $0.510 $ \\
& & & Recall & $0.673$ & $ 0.510 $ \\
& & & F1 & $0.237$ & $\textbf{0.440}$ \\
\midrule
\multirow{12}{*}{Llama-3.3-70b} & \multirow{6}{*}{Trivial} & \multirow{3}{*}{Line-level} & Precision & $ 0.054$ & $0.520 $ \\
& & & Recall & $0.980$ & $0.560 $ \\
& & & F1 & $0.102$ & $\textbf{0.460}$ \\
\cmidrule{3-6}.  
& & \multirow{3}{*}{Chunk-level} & Precision & $0.032$ & $0.510 $ \\
& & & Recall & $0.744$ & $0.520$ \\
& & & F1 & $0.062$ & $ \textbf{0.390}$ \\
\cmidrule{2-6} 
& \multirow{6}{*}{Nontrivial} & \multirow{3}{*}{Line-level} & Precision & $0.200$ & $0.530 $ \\
& & & Recall & $0.986$ & $0.550 $ \\
& & & F1 & $0.333$ & $\textbf{0.510}$ \\
\cmidrule{3-6} 
& & \multirow{3}{*}{Chunk-level} & Precision & $0.131$ & $0.500 $ \\
& & & Recall & $0.531$ & $0.500$ \\
& & & F1 & $0.211$ & $\textbf{0.440}$ \\
\midrule
\multirow{12}{*}{Llama-4-Scout} & \multirow{6}{*}{Trivial} & \multirow{3}{*}{Line-level} & Precision & $0.017$ & $0.500 $ \\
& & & Recall & $0.983$ & $0.530$ \\
& & & F1 & $0.034$ & $\textbf{0.430}$ \\
\cmidrule{3-6}. 
& & \multirow{3}{*}{Chunk-level} & Precision & $0.020$ & $0.500 $ \\
& & & Recall & $0.891$ & $ 0.490$ \\
& & & F1 & $0.040$ & $\textbf{0.370}$ \\
\cmidrule{2-6}.  
& \multirow{6}{*}{Nontrivial} & \multirow{3}{*}{Line-level} & Precision & $0.064$ & $0.510 $ \\
& & & Recall & $0.993$ & $0.560$ \\
& & & F1 & $0.121$ & $\textbf{0.430}$ \\
\cmidrule{3-6} 
& & \multirow{3}{*}{Chunk-level} & Precision & $0.080$ & $0.510 $ \\
& & & Recall & $0.778$ & $ 0.540 $ \\
& & & F1 & $0.145$ & $\textbf{0.410}$ \\
\bottomrule
\end{tabular}
\caption{AI-generated code detection results for HybridCodeAuthorship dataset including model-specific and dataset split results}
\label{tab:exp_results}
\end{table*}

While HybridCodeAuthorship is intended to benchmark the performance of AI-generated code detection approaches at the line-level, for completeness we also report the results of chunk-level experiments. Chunk-level detection involves concatenating consecutive lines of code with the same author, either human or AI. We benchmarked the task of both line- and chunk-level AI-generated code detection on the HybridCodeAuthorship dataset using two state-of-the-art approaches to automated AI-generated code detection.  The first, DroidDetect \cite{orel2025droid} \footnote{\url{https://huggingface.co/project-droid/DroidDetect-Large-Binary}} is a model-based AI-generated code detector that uses ModernBERT \cite{warner2025smarter} as a base-model and is fine-tuned on a collection of human- and AI-generated code. The second,  AIGCode Detector \cite{xu2024detecting},  is a modified version of the DetectGPT \cite{mitchell2023detectgpt} algorithm, which introduced a perturbation-based approach to AI-generated text detection: the target text is perturbed using another pre-trained model, such as T5, and the log probability of the perturbed and non-perturbed variants of the text are compared. This approach leverages the observation that machine-generated text tends to have a lower log probability after minor perturbations, whereas human-authored text is less sensitive to these changes.  AIGCode Detector's modifications include the use of 
CodeBERT \cite{feng2020codebert}, rather than T5,  to perturb the target code, with the proportion of targeted code determined by the perplexity score  of the target code rather than the fixed  percentage utilized by DetectGPT. Additionally,  rather than using the ``perplexity disparity'' between perturbed and unperturbed code to assign a human vs. AI score, a composite score is calculated which combines the perplexity of the target code,  the standard deviation of the perplexity scores calculated over multiple perturbed versions of the code, and  a burstiness score.

F1 scores for the best-performing algorithm relative to the LLM, dataset split (trivial vs nontrivial) and granularity (line- vs chunk-level) have been boldfaced in Table \ref{tab:exp_results}.   AIGCode Detector easily out-performed DroidDetect  across  all experiment variants. Chunk-level detection emerged as the more challenging task which aligns with the original DetectGPT authors' \cite{mitchell2023detectgpt} observation of a  degradation in DetectGPT's performance as a function of text length. As expected, performance on the task of detection in trivial segments, which contain minimal semantic content, was significantly worse than that of nontrivial segments for all variants, with the exception of GPT-OSS-120b.

\section{Limitations}
Although HybridCodeAuthorship is currently the only available dataset that simulates hybrid AI and human authorship at the fine-grained line level, there are some limitations to discuss. 

First, the code testing phase did not succeed in perfectly handling the incompatibility of all legacy code files in CodeSearchNet, which led to a number of false negatives in unit test results. 

Second, the code interleaving phase faced challenges in code generation. Some code samples were not successfully processed by all three LLMs, which resulted in fewer samples in the final benchmark. The challenges mainly came from the finite context length and inconsistent instruction compliance during prompting. There are observable characteristics that are strongly correlated with pipeline failure (for example, code files longer than the LLM context window will always fail). This could impact the real-world representativeness of the sample of code files that make it into HybridCodeAuthorship from CodeSearchNet. Using more advanced LLMs could resolve some of these challenges in future work. Additionally, because we needed to be sure that no AI code was present in the original code we interleaved, we selected CodeSearchNet, which contains code that is 6 years old or older. As a result, HybridCodeAuthorship does not contain recently developed or popularized code libraries.

\section{Conclusion}

In this paper, we present a novel benchmark dataset that will allow the development of algorithms to distinguish AI-generated versus human-authored lines of code, built by a novel data processing pipeline that could be reused for expansion of HybridCodeAuthorship. Future work could expand HybridCodeAuthorship to include other popular programming languages such as Java, JavaScript, and Go. This would ensure greater representativeness across the landscape of software development. Another enhancement could be adding more examples from additional LLMs, which may offer greater diversity of AI-generated coding patterns and greater generalizability of benchmark results. Using LLMs that specialize in code generation and handle longer context than the three LLMs used could lead to more successfully generated interleaved results. Looking ahead, this line of work opens several promising avenues for future research. Future efforts could focus on designing algorithms specifically for the line-level attribution task, possibly by integrating features from code stylometry and other related work.

\nocite{*}
\section{Bibliographical References}\label{sec:reference}
\bibliographystyle{lrec2026-natbib}
\bibliography{references}

\appendix
\vspace{-1em}
\section{Test Error Analysis}
\label{appendix_a}
\begin{figure}[htbp] 
\centering 
\includegraphics[width=\linewidth]{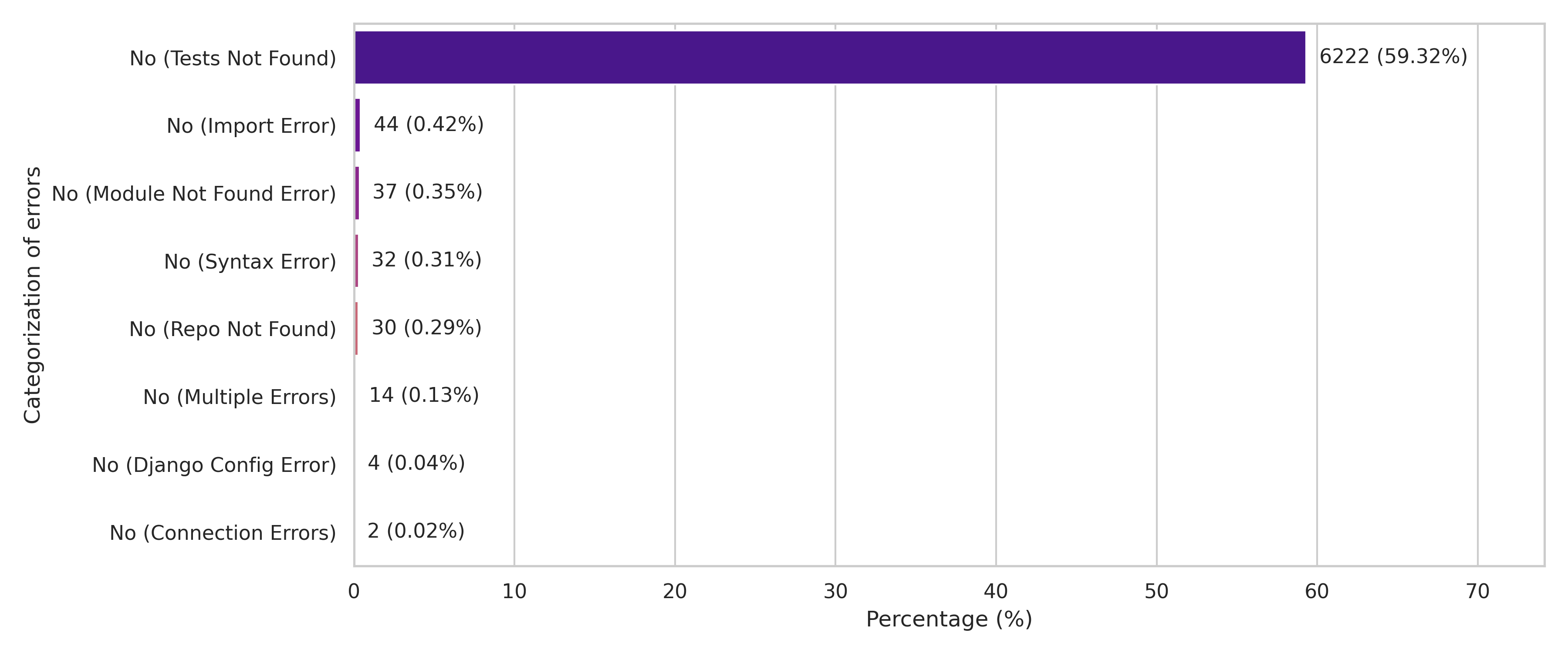}
\caption{Error distribution in testing human-authored code.}
\label{fig:error_human}
\end{figure}

\begin{figure}[htbp] 
\centering 
\includegraphics[width=\linewidth]{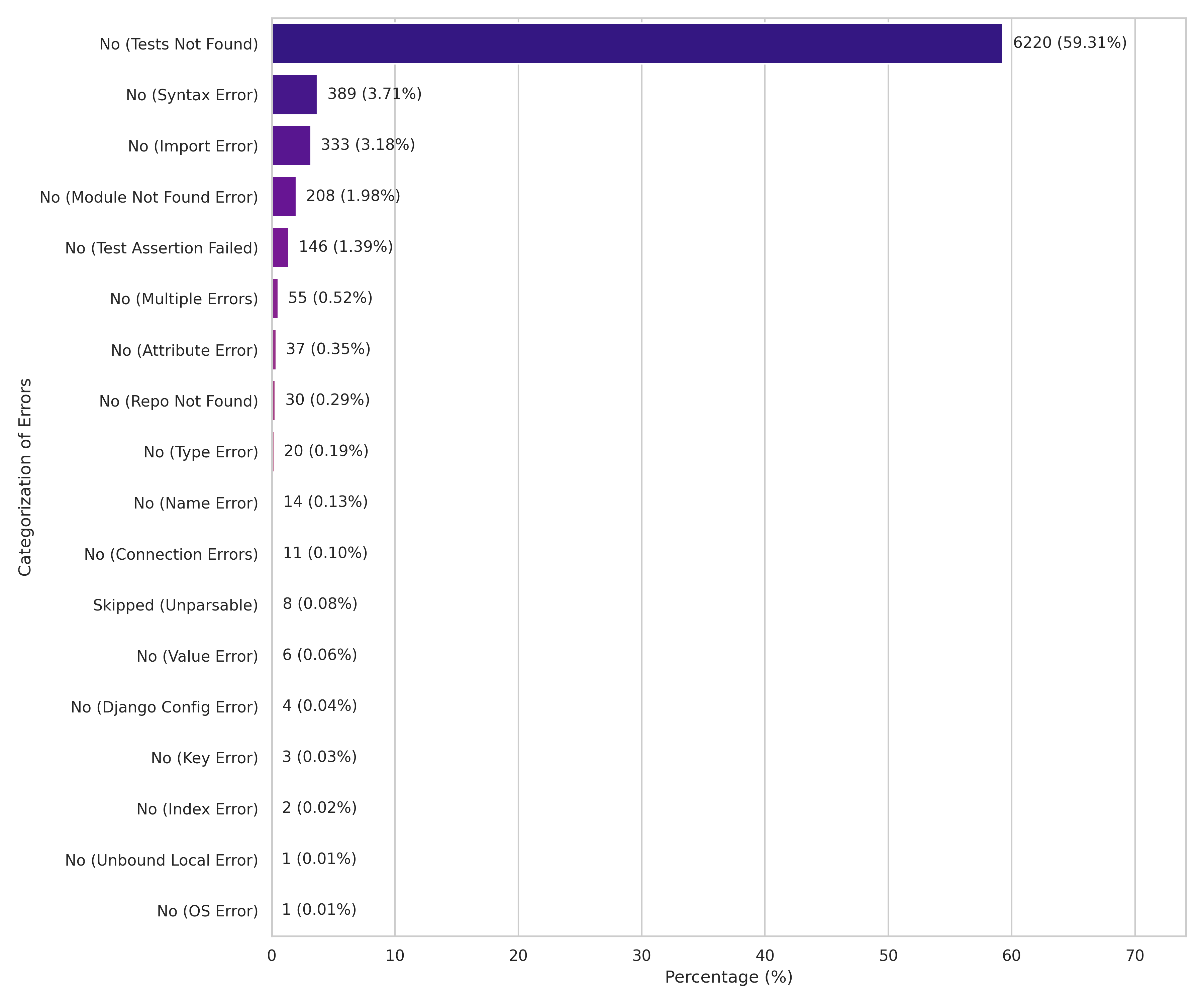}
\caption{Error distribution in testing AI-modified code.}
\label{fig:error_ai}
\end{figure}

We categorized different errors raised in testing both human-authored and AI-modified code. The error distributions for the two scenarios are visualized in Figure \ref{fig:error_human} and Figure \ref{fig:error_ai}, respectively. Both distributions share the most common errors as listed in Section 3.2.4, Test Results Handling. Specific error definitions include \texttt{Multiple Errors} (two or more concurrent errors), \texttt{Django Config Error} (\texttt{ImproperlyConfigured} exceptions), and \texttt{Connection Error} (HTTP/API access failures). Furthermore, AI-modified code by the open-source LLMs in our case tended to produce new errors not observed with human-authored code: \texttt{Attribute Error}, \texttt{Type Error}, \texttt{Name Error} and \texttt{Value Error} are the top errors unique to AI-modified code while the percentage of \texttt{Syntax Error} is magnified. Notably, this observation coincide with the findings in previous literature \citep{rahman2025beyond, tambon2025bugs, wen2024fixing, yin2023natural, jimenez2023swe}.

\section{Code Marking Prompt}
\label{appendix_b}
This is the prompt fed to the LLM for the \textbf{code identification and marking tasks}. 

\begin{quote}
\small\ttfamily
\textbf{Code Identification and Marking Prompt} \\
\rule{\linewidth}{0.4pt} \\
You will receive a code snippet. Your task is to remove some parts of the code, and replace it with placeholder comments that describe what the section you removed must do. 
    
Some criteria for replacement: 
\begin{enumerate}
    \item the sections of code must be clearly atomic and logically separable from surrounding code, such that the selected code can be cleanly described (for example, don't stop the deleted section in the middle of a for loop). 
    \item clearly distinguish comments added in this manner from other comments in the code by starting your comment with ``GENERATE CODE:''
    \item Your comments need to be detailed enough that someone without reference to the original code could plausibly reconstruct the code section with similar functionality (though not necessarily with the same exact syntax)
    \item Try to be not super specific about the exact implementation, the goal is not to describe it so specifically that there is only one possible code string that meets the specified requirements
    \item Do not replace non-significant, non-meaningful lines of code. Examples of these are:
        \begin{itemize}
            \item Blank lines
            \item Comment-only lines
            \item Docstrings
            \item Simple print statements
            \item Simple brackets/parentheses
            \item Simple pass statements
            \item Setting variables to some string, bool or int literal
        \end{itemize}
\end{enumerate}

Remember, you are to REMOVE some parts of the code, and REPLACE the removed code with your comments. So the file may not be a valid, executable code file, and that's ok.
Return nothing else besides the edited code file. You'll also receive the approximate percentage of significant, meaningful lines of code that should be replaced in this manner.

[Approximate Percentage Inserted] \newline
[Code File Inserted]
\end{quote}

\section{Code Writing Prompt}
\label{appendix_c}
This is the prompt fed to the LLM for the \textbf{code writing task}.
 
\begin{quote}
\small\ttfamily
\textbf{Code Writing Prompt} \\
\rule{\linewidth}{0.4pt} \\
We are creating code snippets that have human-written and AI-written code mixed together. 
You are responsible for creating the AI-written portions of the code snippets.  
I will give you a completely human-written code snippet, with some sections removed. the functionality of the removed sections
are described in inline comments that start with ``GENERATE CODE:``. Your task is to create code that meets the requirements described in these comments. 

Return the completed version of the entire code file, including both the original contributions and your contributions. 
Remove all ``GENERATE CODE:'' comments from your output, but preserve other comments. Return nothing else besides this complete code file.

[Code File Inserted]
\end{quote}

\end{document}